\pdfoutput=1
\documentclass[prl,twocolumn,%
               superscriptaddress,%
               amsfonts,amssymb,amsmath,%
               showpacs,%
               titlepage,reprint,
               ]{revtex4}

\usepackage[breaklinks=true]{hyperref}
\usepackage{latexsym}
\usepackage{verbatim} 
\usepackage{natbib}
\usepackage{graphicx}
\usepackage{textcomp}
\usepackage{xcolor}
\usepackage{xspace} 
\usepackage{ulem} 
\newcommand{\pgi}{Peter Gr\"{u}nberg Institut (PGI-3), Forschungszentrum J\"{u}lich, 52425 J\"{u}lich, Germany}
\newcommand{\jara}{J\"{u}lich Aachen Research Alliance (JARA), Fundamentals of Future Information Technology, 52425 J\"{u}lich, Germany}
\newcommand{\fhi}{Fritz-Haber-Institut der Max-Planck-Gesellschaft, 14195 Berlin, Germany}
\newcommand{\mpi}{Max Plank Institute for Solid Research, Heisenbergstra{\ss}e, 70569 Stuttgart, Germany}
\newcommand{\diam}{Diamond Light Source Ltd, Didcot, OX110DE, Oxfordshire, United Kingdom}
\newcommand{\dme}{Department of Mechanical Engineering and Material Science, Duke University, Durham, NC 27708 USA}
\newcommand{\etal}{\textit{et al.}\xspace}

\begin{document}
\title{Approaching ideal graphene: The structure of hydrogen-intercalated graphene on 6H-SiC(0001)}
\author{J. Sforzini}
\affiliation{\pgi}
\affiliation{\jara}

\author{L. Nemec}
\affiliation{\fhi}

\author{T. Denig}
\affiliation{\mpi}

\author{B. Stadtm\"{u}ller\footnote{Present address: Department of Physics and Research Center OPTIMAS, University of Kaiserslautern, Erwin-Schr\"{o}dinger-Strasse 46, 67663 Kaiserslautern, Germany}}
\affiliation{\pgi}
\affiliation{\jara}

\author{T.-L. Lee}
\affiliation{\diam}

\author{C. Kumpf}
\affiliation{\pgi}
\affiliation{\jara}

\author{S. Soubatch}
\affiliation{\pgi}
\affiliation{\jara}

\author{U. Starke}
\affiliation{\mpi}

\author{P. Rinke}
\affiliation{\fhi}
\affiliation{COMP/Department of Applied Physics, Aalto University, P.O. Box 11100, Aalto FI-00076, Finland}

\author{V. Blum}
\affiliation{\fhi}
\affiliation{\dme}

\author{F.C. Bocquet}
\email{f.bocquet@fz-juelich.de}
\affiliation{\pgi}
\affiliation{\jara}

\author{F.S. Tautz}
\affiliation{\pgi}
\affiliation{\jara}

\date{\today}

\begin{abstract}
We measure the adsorption height of hydrogen-intercalated quasi-free-standing monolayer graphene on the (0001) face of 6H silicon carbide by the normal incidence x-ray standing wave technique. A density functional calculation for the full ($6 \sqrt{3} \times 6 \sqrt{3}$)-R30$^\circ$ unit cell, based on a van der Waals corrected exchange correlation functional, finds a purely physisorptive adsorption height in excellent agreement with experiments, a very low buckling of the graphene layer, a very homogeneous electron density at the interface and the lowest known adsorption energy per atom for graphene on any substrate. A structural comparison to other graphenes suggests that hydrogen intercalated graphene on 6H-SiC(0001) approaches ideal graphene.
\pacs{73.20.Hb, 61.48.Gh, 68.49.Uv, 71.15.Mb}
\keywords{quasi-free-standing monolayer graphene, hydrogen intercalation, XSW, DFT}
\end{abstract}

\maketitle

During the last decade, graphene attracted a broad interest for its structural and electronic properties \cite{zhang2006,katsnelson2006} which makes it a promising material for a wide range of applications, e.g., transistors in nanoscale devices \cite{Fiori2013} and energy storage \cite{Wang2010b}. The exact material properties of graphene depend on the growth conditions on a given substrate and its interaction with the substrate. In order to maintain its unique electronic properties, it is important to understand the coupling between the graphene layer and the substrate, in terms of covalent and non-covalent bonding, residual corrugation and doping.

Large-scale ordered epitaxial graphene can be grown on various metal substrates. However, the metallic contact to the graphene layer determines its transport properties through, for instance, buckling or doping of the graphene layer \cite{Busse2011, Gamo1997}. It is therefore paramount to find a substrate for which the interactions are minimized in order to preserve the extraordinary properties of a single graphene layer. In addition, the use of a non-metallic substrate is necessary to be able to use graphene, for instance, in electronic devices.

In this context, graphene growth on various faces of the wide band-gap semiconductor silicon carbide (SiC) appears appealing. Riedl \etal \cite{Riedl2009} demonstrated the possibility to decouple graphene from SiC by intercalation of hydrogen atoms (quasi-free-standing monolayer graphene or QFMLG). It is known from the band structure and core-levels of graphene \cite{Forti2014a} that the intercalation process reduces the interaction with the substrate substantially (removal of covalent bonds and less doping). However, these measurements are indirect and, moreover, for weakly interacting graphenes the sensitivity of angle-resolved photoelectron spectroscopy (ARPES) becomes insufficient to assess the interaction with the substrate \cite{Johannsen2013}.

An alternative criterion to gauge the interaction strength of graphene with a substrate is its adsorption height. However, for hydrogen-intercalated graphene, the adsorption height is not known experimentally. Moreover, it is not clear whether for such a weakly interacting system this height can be calculated reliably as it is entirely determined by van der Waals (vdW) interactions, which are difficult to treat. In this letter, we present a density functional theory (DFT) calculation of QFMLG using the full unit cell and a vdW correction to the exchange correlation potential in which the dispersion coefficients are derived from the self-consistent electron density \cite{tkatchenko2009amv}. The calculation yields an adsorption height that is indicative of a purely vdW interaction. We validate this calculation with an accurate experimental height determination by normal incidence x-ray standing wave (NIXSW) and find an excellent agreement. By comparing our results to the adsorption height of graphene on various substrates taken from the literature, we demonstrate that QFMLG on SiC has the least graphene-substrate interaction among all studied systems. This is confirmed in our DFT calculations by a very low buckling of the graphene layer, a very homogeneous electron density at the interface and the lowest known adsorption energy per atom for graphene on any substrates.

Experiments and calculations were carried out for graphene on 6H-SiC(0001). Due to its smoothness and homogeneity, the Si-terminated surface of the 6H-SiC is widely used to achieve a controlled formation of high quality epitaxial graphene monolayers \cite{forbeaux1998hgo, Emtsev2009, starke2009ego, riedl2010rev}. However, the first honeycomb carbon layer formed on the SiC(0001) surface consists of $\textsl{sp$^2$}$ and $\textsl{sp$^3$}$ hybridized orbitals leading to the formation of the so-called zerolayer graphene (ZLG) \cite{Varchon2007,Emtsev2008}. Since some of its atoms are covalently bonded to the Si atoms of the SiC surface, the ZLG does not show the typical Dirac cone in its band structure \cite{Forti2014a}. To recover the typical electronic properties  of graphene, namely linear dispersion of the  $\pi $ and $ \pi^*$ bands at the K point of the hexagonal Brillouin zone, the formation of an additional graphene layer on top of the ZLG is required, generating an epitaxial monolayer graphene (EMLG). Although the ZLG decouples the EMLG from the substrate, it is still considered to be a main obstacle for the development of graphene-based electronic devices because of the residual interactions. In fact, the Si dangling bonds in the top layer induce a significant doping in the EMLG even through the ZLG \cite{Bostwick2007}. Replacing the carbon ZLG by a more passivating layer is therefore necessary to produce free standing-like graphene on the SiC substrate. This can be achieved by hydrogen intercalation. The hydrogen atoms passivate the Si atoms in the top SiC bilayer. In this process the bonds between the ZLG and Si atoms are broken, the $\textsl{sp$^3$}$ atoms in the ZLG re-hybridize, and the ZLG is lifted above the hydrogen atoms at the interface, forming QFMLG. Thus, hydrogen takes over the decoupling role of the ZLG layer in the EMLG.

\begin{figure}
  \centering
  \includegraphics [width=\columnwidth]{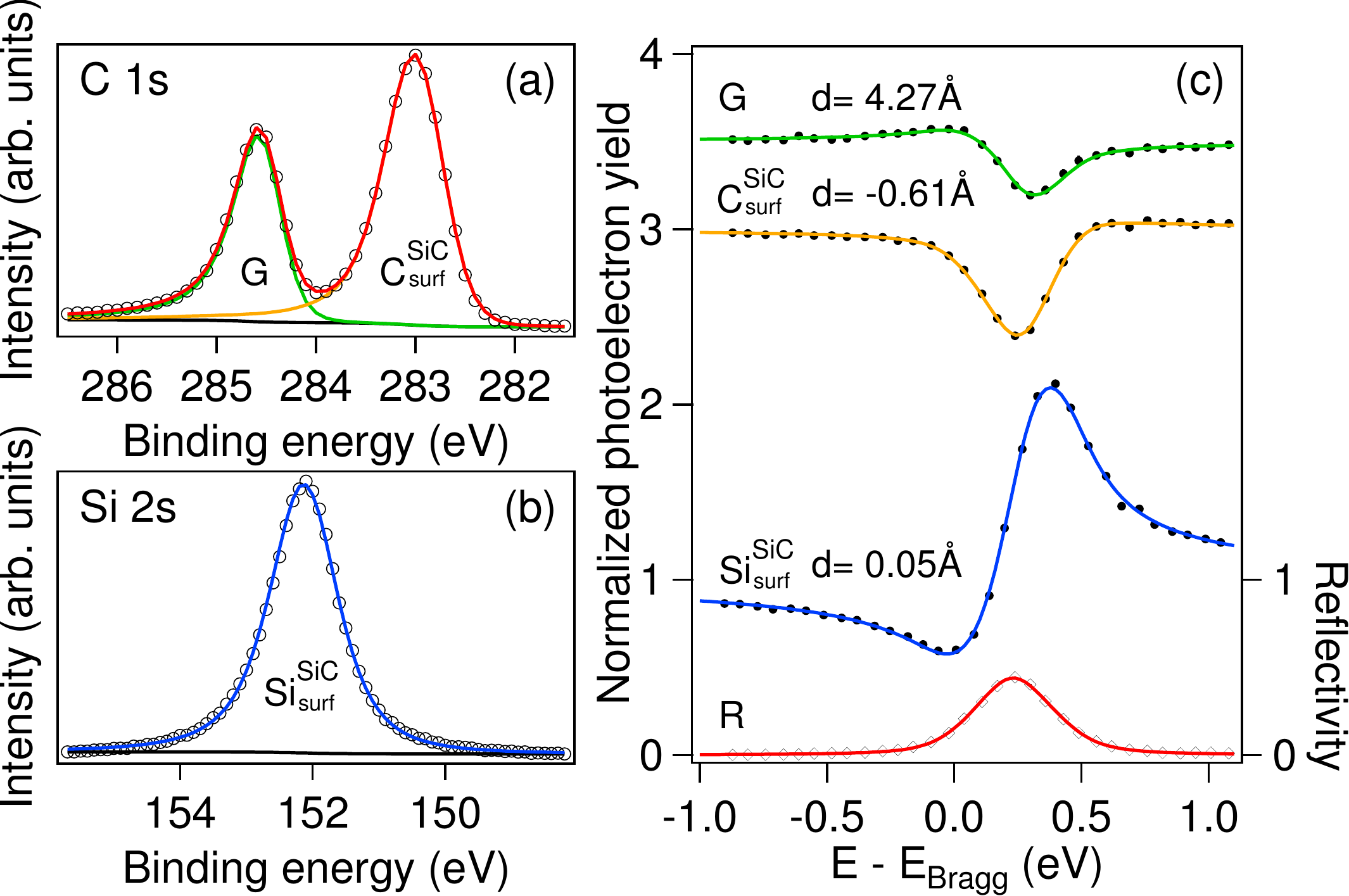}
  \caption{NIXSW data measured for QFMLG on 6H-SiC(0001). (a) C~$1s$ core-level, fitted with two asymmetric Lorentzians. G and $\mathrm{C_{surf}^{SiC}}$ correspond to the graphene and the surface carbon atoms of SiC, respectively.  (b) Si~$2p$ core-level fitted with a pseudo-Voigt function. Both were measured with a photon energy of 2494~eV. (c) Black dots: experimental photoelectron yield curves vs. photon energy relative to the (0006) Bragg energy (2463~eV). The error bars, estimated according to \cite{Mercurio2013}, are smaller than the symbols. Fits to the yields curves for the surface atoms of SiC ($\mathrm{Si_{surf}^{SiC}}$, $\mathrm{C_{surf}^{SiC}}$) and graphene (G) are shown in blue, orange and green, respectively \cite{suppl, Woodruff1998, Woodruff2005, Vartanyants2001}. The reflectivity R is plotted with black diamonds and its best fit in red. The absolute distances for each component are given with respect to the bulk-extrapolated silicon planes. The error bar for each value is $\pm0.04$~{\AA}.}
  \label{fig1}
\end{figure}

The NIXSW experiments were performed in an ultra high vacuum end-station at the I09 beamline at Diamond Light Source (Didcot, United Kingdom) equipped with a VG Scienta EW4000 hemispherical electron analyzer (acceptance angle of $60^\circ$) perpendicular to the incident beam direction. All data sets were recorded at room temperature and in a normal incidence geometry. A photon energy of approximately 2463~eV was used to excite the 6H-SiC(0006) reflection, which has a Bragg plane spacing of 2.517~{\AA}. The NIXSW method, combining dynamical x-ray diffraction and photoelectron spectroscopy, is a powerful tool for determining the vertical adsorption distances at surfaces with sub-{\AA} accuracy and high chemical sensitivity. The samples were prepared by thermal decomposition of SiC to produce the ZLG and then by annealing up to $700^\circ$C in molecular hydrogen at atmospheric pressure to produce the QFMLG. After being transported in air to the beamline, the samples were outgassed in the end-station before the x-ray measurements. ARPES using monochromatized He I$\alpha$ radiation and low energy electron diffraction (LEED), shown in \cite{suppl}, were used to check the electronic and structural properties. The x-ray results were obtained from two samples at different spots and showed no sample and position dependence.

\begin{figure*}[t]
  \centering
  \includegraphics[width=.9\textwidth]{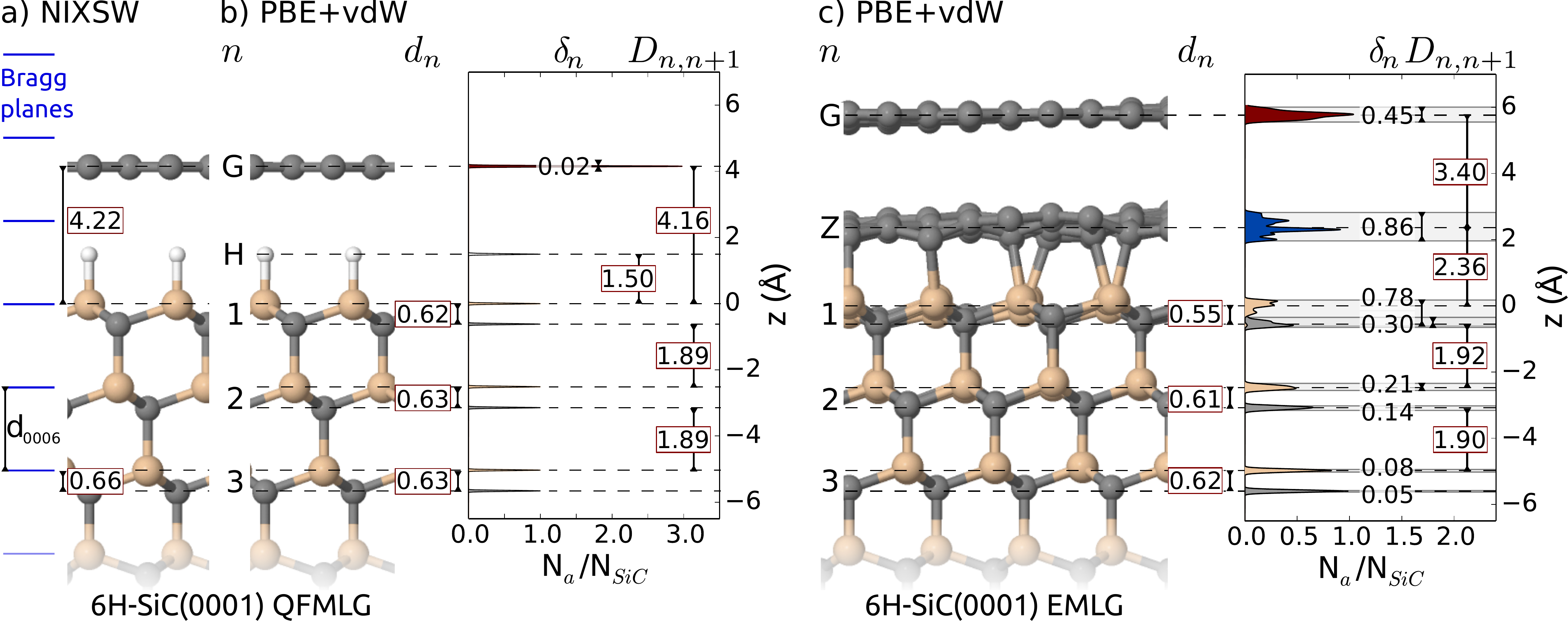}
  \caption{\label{Fig:histograms} (a) Vertical distances measured by NIXSW on QFMLG. The position of the Bragg planes around the surface are indicated by blue lines. PBE+vdW calculated geometry for (b) QFMLG and for (c) EMLG on 6H-SiC(0001) and histograms of the number of atoms $\mathrm{N}_{a}$ vs. the atomic coordinates (z) relative to the topmost Si layer (Gaussian broadening: $0.02$~{\AA}). $\mathrm{N}_{a}$ is normalized by $\mathrm{N}_{SiC}$, the number of SiC unit cells. $D_{n,n+1}$ is the distance between the layer $n$ and $n+1$, $d_{n}$ gives the Si-C distance within the SiC bilayer $n$, and $\delta_{n}$ the corrugation of the layer $n$. All values are given in {\AA}.}
\end{figure*}

The surface SiC ($\mathrm{C_{surf}^{SiC}}$) and graphene components of the C~$1s$ spectrum are found at binding energies of 283.1~eV and 284.7~eV, respectively (Fig.~\ref{fig1}(a)), and the Si~$2s$ is found at 152.2~eV (Fig.~\ref{fig1}(b)) for the surface SiC ($\text{Si}_{\text{surf}}^{\text{SiC}}$). The photoelectron yield of each chemical species is deduced from the peak area determined by a line-shape analysis of the core-level spectrum. This is repeated for all photon energy steps over a 2~eV range around the Bragg energy ($\mathrm{E_{Bragg}}$) for all three species. Following a well established procedure \cite{suppl}, we fit the final reflectivity and photoelectron yield curves with dynamical diffraction theory to determine the heights of the three different species with respect to the bulk-extrapolated SiC(0006) atomic plane. The $\mathrm{C_{surf}^{SiC}}$ atoms are located at $0.61\pm0.04$~{\AA} below the bulk-extrapolated silicon plane and the $\mathrm{Si_{surf}^{SiC}}$ $0.05\pm0.04$~{\AA} above. Thus we obtain an experimental Si-C distance of $0.66\pm0.06$~{\AA}, in agreement with the SiC crystalline structure \cite{lee2010a, Wang2010a}. In the same way, we find the adsorption height of the graphene layer with respect to the topmost Si layer to be $4.22\pm0.06$~{\AA}, as shown in Fig.~\ref{Fig:histograms}~(a). We note that this height is approximately equal to the sum of the vdW radii of carbon and hydrogen (plus the Si-H distance of approximately 1.50~\AA), and thus indicates the absence of interactions besides vdW.

To test whether the structure of this predominantly vdW interacting interface can be predicted using DFT and to gain a detailed understanding of how hydrogen decouples the graphene layer from the substrate, we performed DFT calculations for the QFMLG and EMLG. The calculations were carried out using the all-electron, localized basis set code FHI-aims (‘tight’ settings) \cite{blum2009aim, havu2009eoi, auckenthaler2011pso, Ren2012} and the Perdew-Burke-Ernzerhof (PBE) functional \cite{perdew1997gga} with a correction for vdW effects (PBE+vdW). There are many different approaches to include long-range dispersion effects into DFT calculations \cite{dobson2012cod}. We use the well established Tkatchenko-Scheffler \cite{tkatchenko2009amv} method to efficiently include vdW effects into large-scale DFT calculations with thousands of atoms. It is a pairwise approach, where the effective C6 dispersion coefficients are derived from the self-consistent electron density. For the bulk lattice parameter of the 6H-SiC polytype we found $a= 3.082$~{\AA} and $c=15.107$~{\AA}. We stress that we investigate the QFMLG and EMLG reconstructions in the experimentally observed, large commensurate ($6 \sqrt{3} \times 6 \sqrt{3}$)-R30$^\circ$ supercell consisting of 6 SiC-bilayers under each surface reconstruction (1850 and 2080 atoms for the QFMLG and EMLG, respectively). We fully relaxed the top three SiC bilayers and all planes above (residual energy gradients $< 8 \cdot 10^{-3}$~eV/{\AA}).

Figure~\ref{Fig:histograms} compares the measureed structure of QFMLG (Fig.~\ref{Fig:histograms}~(a)) and the calculated structure of the QFMLG and EMLG (Fig.~\ref{Fig:histograms}~(b,c)) on 6H-SiC predicted at the PBE+vdW level. In addition, we include a histogram of the atomic $z$ coordinates relative to the top Si layer normalized by the number of SiC unit cells. For illustration purposes, we broadened the histogram lines using a Gaussian with a width of $0.02$~{\AA}. For the QFMLG, we found a bulk-like distance of $1.89$~{\AA} between the SiC bilayers. The Si-C distance within the top SiC bilayer ($0.62$~{\AA}) and the remaining Si-C bilayer distances are practically bulk-like ($0.63$~{\AA}), in good agreement with the experimental result ($0.66\pm0.06$~{\AA}). The distance between the top Si-layer and the graphene layer is $4.16$~{\AA} for 6H-SiC, again in good agreement with the measured $4.22\pm0.06$~{\AA}. The $0.02$~{\AA} corrugation of the graphene layer is very small. For the hydrogen layer and all layers underneath the corrugation is $<10^{-2}$~{\AA}.

\begin{figure}
  \centering
  \includegraphics [width=.8\columnwidth]{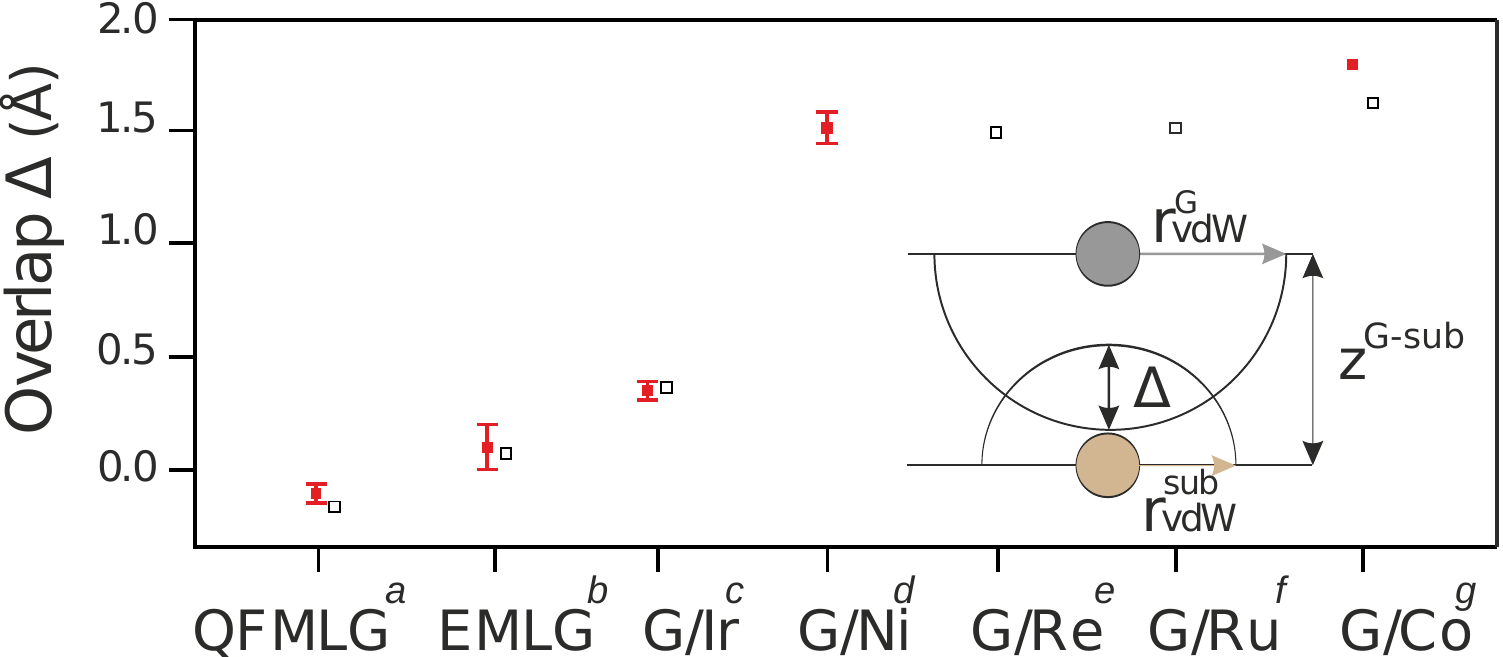}
  \caption{Comparison of the overlap $\Delta$ for different epitaxial graphene systems. $\Delta$ is calculated by subtracting $z^{\mathrm{G-sub}}$ from the sum of graphene and substrate vdW radii. The empty and the filled squares correspond to DFT and measured values, respectively. \textit{$^a$}: this work, \textit{$^b$}: \cite{Emery2013, nemec2013tec}, \textit{$^c$}: \cite{Busse2011}, \textit{$^d$}: \cite{Gamo1997}, \textit{$^e$}:\cite{Miniussi2011}, \textit{$^f$}: \cite{Wang2010}, \textit{$^g$}: \cite{eom2009}.}\label{Fig4}
\end{figure}

The situation is very different for the EMLG in Fig.~\ref{Fig:histograms}~(c). Here a significant buckling of the graphene layer is observed \cite{goler2013, Emery2013}. In the EMLG, the interface between bulk SiC and graphene is formed by the partially covalently bonded ZLG. This interface layer is corrugated by $0.86$~{\AA}, leading to a buckling of the graphene layer of $0.45$~{\AA}, as well as a strong corrugation of $0.78$~{\AA} in the top Si layer. The interlayer distance ($1.92$~{\AA}) between the top substrate bilayers is increased in comparison with the bulk value, while the Si-C distance within the topmost SiC bilayer is substantially reduced, see Fig.~\ref{Fig:histograms}~(c). In summary, our calculations provide a valid description of the graphene SiC interface, as they reproduce quantitatively the NIXSW-measured graphene-Si distances for both QFMLG and EMLG \cite{Emery2013, nemec2013tec}.

Using a smaller approximated ($\sqrt{3} \times \sqrt{3}$)-R30$^\circ$ cell (50 atoms for QFMLG), we tested the influence of the exchange correlation functional and the type of vdW correction on the geometries \cite{suppl}. The Si-graphene distance for QFMLG calculated in the approximated cell using the same methodology as discussed above is $4.25$~{\AA}. When we applied the highest level of theory using the Heyd-Scuseria-Ernzerhof hybrid functional (HSE06) \cite{krukau2006iot} with a vdW correction incorporating many-body effects (HSE06+MBD) \cite{ambrosetti2014lrc, tkatchenko2013mbd, tkatchenko2012aae}, the Si-graphene distance increased slightlty to $4.26$~{\AA}. The difference of $0.01$~{\AA} between PBE+vdW and HSE06+MBD is negligible. We can thus conclude that changes in the predicted vertical structure of the ($6 \sqrt{3} \times 6 \sqrt{3}$)-R30$^\circ$ supercell would also be small even if higher level approximations to the exchange correlation functional were employed.

Comparing the buckling of QFMLG ($0.02$~{\AA}) and of EMLG ($0.45$~{\AA}), we can conclude that QFMLG is much more ideal for device applications than EMLG. This is confirmed by a qualitative analysis in terms of overlapping vdW radii \cite{Bondi1964, Batsanov2001} where the overlap is defined by $\Delta=r_{\mathrm{vdW}}^{\mathrm{G}}+r_{\mathrm{vdW}}^{\mathrm{sub}} -z^{\mathrm{G-sub}}$ with $r_{\mathrm{vdW}}^{\mathrm{G}}$, $r_{\mathrm{vdW}}^{\mathrm{sub}}$ and $z^{\mathrm{G-sub}}$ being the vdW radii of graphene and of the atoms immediately below the graphene layer, and the measured distance between the graphene and the topmost atoms of the substrate, respectively. $\Delta>0$ means that the vdW radii of the graphene and of the substrate overlap, indicating some degree of chemical interaction. On the other hand, for $\Delta \lesssim 0$, the graphene-substrate interaction is expected to be very weak. In Fig.~\ref{Fig4}, the overlap is plotted for QFMLG in comparison with other systems for which the adsorption heights have been measured or calculated. Epitaxial graphenes on SiC exhibit the lowest overlaps and QFMLG has by far the lowest value. This is also reflected in the low adsorption energy calculated for QFMLG, which is 59~meV/atom, significantly smaller than the corresponding values for EMLG (89.2~meV/atom) and graphite (81~meV/atom) \footnote{In ref \cite{Busse2011}, a value of 50~meV/atom has been found for graphene on Ir(111). However this value is an average over a chemically modulated physisorbed graphene layer. Hence, parts of the layer have a much larger adsorption energy.}.

\begin{figure}
 \centering
 \includegraphics[width=.8\columnwidth]{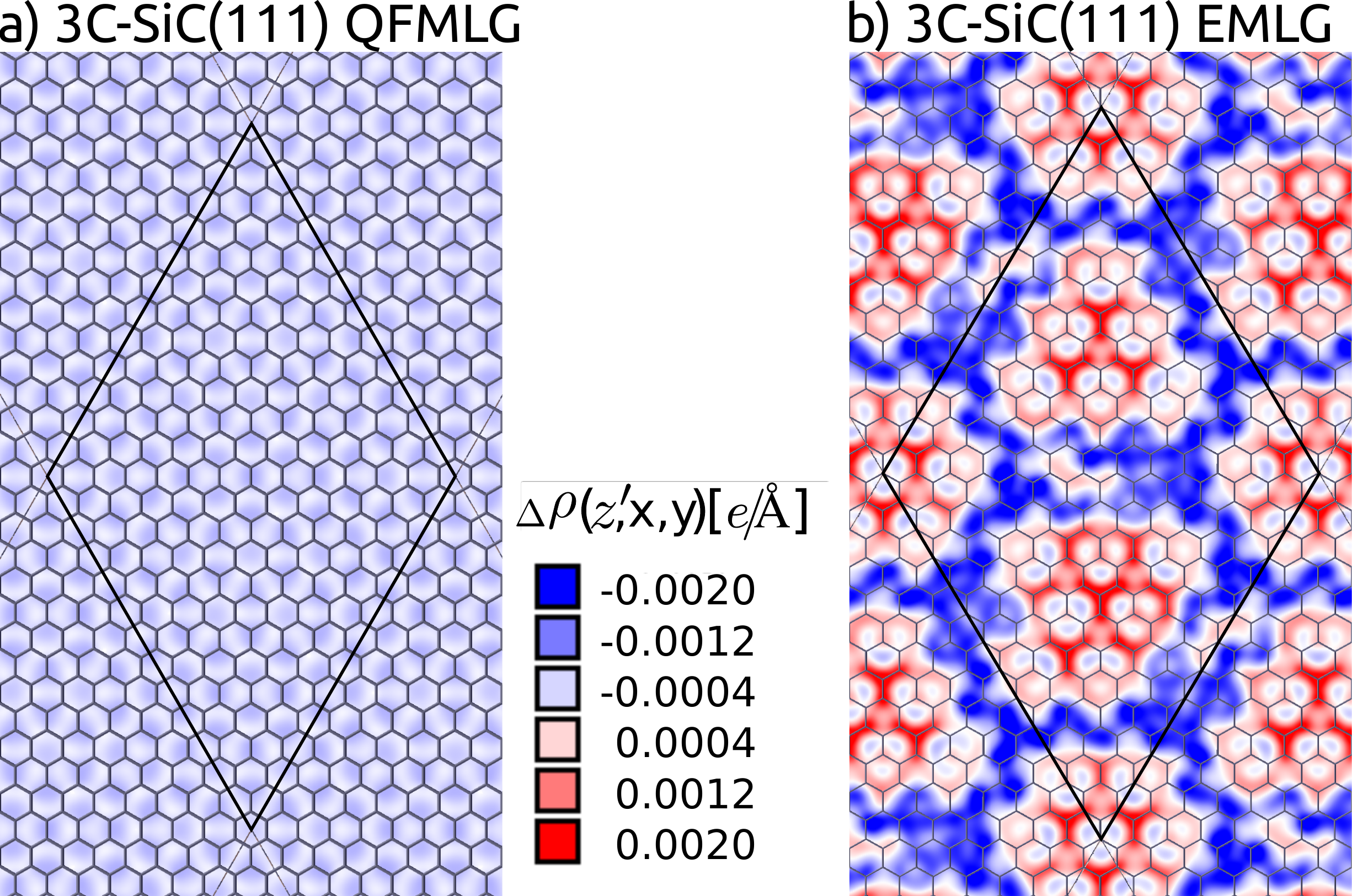}
 \caption{Electron density differences $\Delta \rho(r) = \rho_{\text{full}}(r) - (\rho_{\text{G}}(r) + \rho_{\text{sub}}(r))$ for (a) 3C-SiC QFMLG and (b) 3C-SiC EMLG, calculated in the $x$-$y$-plane between the substrate and the graphene layer (1.08~{\AA} below the graphene layer). The ($6 \sqrt{3} \times 6 \sqrt{3}$)-R30$^\circ$ supercell is shown in black and the graphene layer in gray.}
 \label{fig:dens_map}
\end{figure}

Finally, we show that purely physisorptive adsorption with negligible buckling translates into a more decoupled electronic structure of the graphene. For this purpose, we calculate the change of electron density at the interface for QFMLG and EMLG. The calculations were performed with a 3C-SiC substrate as it allows us to use a smaller substrate thickness (4 layers instead of 6 for 6H-SiC) and renders the calculation more affordable. The SiC polytype (3C and 6H) is known not to influence the surface reconstructions \cite{Schardt2000, Pankratov2012}. We confirmed this by DFT for QFMLG and EMLG on both 6H- and 3C-SiC \cite{suppl}. The electron density of the system is represented on an evenly distributed grid for the full system $\rho_{\text{full}}(r)$. Similarly, the graphene layer $\rho_{\text{G}}(r)$ and the substrate $\rho_{\text{sub}}(r)$, calculated in isolation from each other, include the hydrogen layer for QFMLG and the ZLG for EMLG. The electron density difference $\Delta \rho(r)$ is given by $\Delta \rho(r) = \rho_{\text{full}}(r) - (\rho_{\text{G}}(r) + \rho_{\text{sub}}(r))$. Fig.~\ref{fig:dens_map}  shows $\Delta \rho(r)$ in the $x$-$y$-plane at a given height between the substrate and the graphene layer. The resulting pattern is very similar for any chosen height \cite{suppl}. In QFMLG, all Si atoms are saturated by hydrogen \cite{Bocquet2012} resulting in negligible variations of the charge density within the $x$-$y$-plane  as seen in Fig.~\ref{fig:dens_map}~(a). For EMLG, see Fig.~\ref{fig:dens_map}~(b), the electron density is modulated by the interplay of saturated and unsaturated Si bonds to the ZLG layer. The negligible $\Delta \rho(r)$ of QFMLG is an additional indication for the improved decoupling of the graphene layer from the substrate, thus preventing its buckling. This is in agreement with STM results \cite{goler2013} showing no corrugation within the experimental accuracy.

In conclusion, we have shown that DFT PBE+vdW calculations,  for the large experimentally observed unit cell, accurately predict the adsorption height of QFMLG, in agreement with NIXSW measurements. QFMLG is the system having the largest adsorption distance among studied graphene-substrate systems, in particular, the overlap vanishes, suggesting a very effective decoupling of the graphene layer. Indeed the calculations show that in comparison to EMLG, QFMLG is a very flat graphene layer with a very homogeneous electronic density at the interface. This significant difference translates into a dramatic improvement of transistors after hydrogen intercalation \cite{Hertel2012, Liu2014}. It suggests that the adsorption distance is a valid parameter to assess the ideality of graphene.

\begin{acknowledgments}
F.C.B. acknowledges financial support from the Initiative and Networking Fund of the Helmholtz Association, Postdoc Programme VH-PD-025. This research has been supported by the Academy of Finland through its Centres of Excellence Program (project no. 251748). This work was partially supported by the DFG collaborative research project 951 ``HIOS''. The authors would like to thank S. Schr\"{o}der and C. Henneke for their support during the XSW beam-time, and D. McCue for his excellent technical support at the I09 beam-line.
\end{acknowledgments}

\bibliographystyle{apsrev4-1}
\bibliography{paper_arxiv}
\end {document}